# Quantitative interpretation of binding reactions of rapidly diffusing species using fluorescence recovery after photobleaching

George D.Tsibidis[*]

Institute of Electronic Structure and Laser, Foundation for Research and Technology, P.O.Box 1385, Vassilika Vouton, 71110 Heraklion, Crete, Greece

**Abstract**

Fluorescence recovery after photobleaching (FRAP) measurements offer an important tool for analysing diffusion and binding processes. Confocal Scanning Laser Microscopes (CSLM) that are used in FRAP experiments bleach regions with a radially-Gaussian distributed profile. Previous attempts to derive analytical expressions in the case of processes governed by fast diffusion have overlooked the characteristics of the instruments used to perform FRAP measurements and therefore led to approximating solutions. In the present paper, bleaching laser beam characteristics will be incorporated into an improved model to provide a more rigorous and accurate method. The proposed model simulates binding inside bounded regions and it leads to FRAP curves that depend on the on- and off-rates which can be employed to determine the rate constants. It can be used in conjunction with experimental data acquired with CSLM to investigate the biophysical properties of proteins in living cells. The model aims to improve the accuracy when determining rate constants by taking into account a more realistic scenario of the light-matter interaction.

**Keywords:** FRAP; binding mechanisms; instantaneous processes; intracellular dynamics; mathematical modelling;

[*] Corresponding author, Tel: 00302810391912, Fax: 0030-2810391569

E-mail address: tsibidis@iesl.forth.gr

# 1   Introduction

FRAP has become a very useful tool to measure the translational transport within sub-cellular systems (for a review, see (Carmo-Fonseca *et al.*, 2002, Carrero *et al.*, 2003, Sprague & McNally, 2005)). The technique has been widely applied to systems expressing GFP fusion proteins for the study of a variety of problems including measuring diffusion coefficients of macromolecules in solution (Axelrod *et al.*, 1976, Soumpasis, 1983), probe diffusion in tissues (Berk *et al.*, 1993) and cytoplasm (Luby-Phelps *et al.*, 1987) and finally, describe transport in the endoplasmic reticulum, Golgi (Lippincott-Schwartz *et al.*, 2001) and inside the cell nucleus (Patterson & Lippincott-Schwartz, 2002).

The underlying theoretical framework for most models that aim to investigate intracellular processes and interactions in infinite or bounded domains is a set of reaction-diffusion equations (Carrero et al., 2003, Hinow *et al.*, 2006, Carrero *et al.*, 2004a, Tardy *et al.*, 1995, Carrero *et al.*, 2004b, Sprague *et al.*, 2006, Sprague *et al.*, 2004, Beaudouin *et al.*, 2006, Braga *et al.*, 2007). The average time for diffusion between binding events, the average residency time of proteins in bound form and the relative values of the binding and unbinding rate constants are the parameters that can determine regions where simplistic behaviours hold (Braga et al., 2007, Sprague et al., 2006, Sprague et al., 2004). Basically, discussing complex combinations of intracellular transport together with binding-unbinding events requires one to distinguish between various scenarios. For convenience, one can focus on three boundary situations (Sprague et al., 2004, Tsibidis & Ripoll, 2008): (i) Diffusion is slow in comparison to the binding-unbinding events and thus dominates dynamics, (ii) Diffusion and binding-unbinding take place on comparable timescales, thus leading to complex transport between traps, (iii) Diffusion is fast compared to the binding-unbinding equilibrium, such that the net-process is governed by the binding-unbinding reaction. The present paper focuses on the third case, which is often encountered in the ATP depletion in cells via sodium azide treatment (Sprague et al., 2004) and the binding of nuclear GFP-actin (Carrero et al., 2004a, McDonald *et al.*, 2006). One critical feature that is overlooked from consideration in the derivation of the analytical expression is the form of the bleaching profile (Sprague et al., 2006, Sprague et al., 2004). For simplicity, most models assume that the laser bleaching beam produces a circularly uniform profile. Previous studies have shown, nevertheless, that the bleaching profile obtained on a CSLM is more closely modelled by a Gaussian (in the radial and axial direction), provided that the bleach depth is sufficiently shallow (Braga *et al.*, 2004, Braeckmans *et al.*, 2003, Tsibidis & Ripoll, 2008). Thus, a step-function type bleaching profile leads to an overestimation of the fluorescence depleted area. Although, the form of the bleaching profile will not affect the reaction-diffusion of the molecules, it will influence the fluorescence recovery curve. As a result, the incorporation of a more precise bleaching profile into any theoretical model would lead to a correction in the estimation of the parameters that characterise molecular transport (Tsibidis & Ripoll, 2008).

In the present paper, we will formulate a method to derive an analytical expression that generates the fluorescence recovery curve for processes characterised by instantaneous diffusion. Throughout this work, the bleaching profile will be assumed to be radially-Gaussian. A simulation to test the dependence of fluorescence redistribution on parameters related to the laser beam characteristics will be incorporated into our approach and a comparison to results obtained by employing previous models will be conducted. The model will be validated with a set of experimental data derived from the study of binding-unbinding events of nuclear proteins. We believe that our analysis can further provide an efficient tool for fitting experimental data that will lead to obtain significant insight into the quantitative characterisation of any type of species that diffuse rapidly and react with particular components in bounded domains.

## 2  Model formulation

In FRAP experiments, a fluorescent region is irradiated with a high intensity focused laser beam that causes a 'bleaching' which is described by an irreversible first-order reaction. The bleaching Point Spread Function (PSF), $I_{bleach}(r)$, is considered to be radially Gaussian distributed according to the formula

$$I_{bleach}(r) \propto \exp(-2r^2/r_o^2) \qquad (1)$$

where $r_o$, the radial resolution, denotes the half-width of the laser beam at $e^{-2}$ intensity (Axelrod et al., 1976) along the $r$-axis. The radial resolution is related to the actual profile that penetrates a cylindrical volume in the specimen. Throughout this work, it will be assumed that a FRAP experiment is conducted on a bounded region that is approximated with a cylinder of radius $b$ and height $H$ (Fig.1). The region contains molecules that diffuse and interact with immobile components inside the domain. To observe the molecular kinetics, all molecules have been fused with fluorescent markers. To a good approximation, if a region is photobleached using a low numerical aperture (NA), the resulting fluorescence depletion is the same on average in every specimen's plane (Tsibidis & Ripoll, 2008).

The instantaneous concentration of the fluorescent material immediately after bleaching (at $t=0$) is given by the following expression (Braga et al., 2004, Blonk *et al.*, 1993, Axelrod et al., 1976)

$$C_{unbleached}(r, t=0) = C_i \exp(-K \exp(-2r^2/r_o^2)) = C_i \sum_{n=0}^{\infty} \frac{(-K)^n}{n!} \exp(-2nr^2/r_o^2) \qquad (2)$$

where $K$ denotes the bleach constant and $C_i$ represents the pre-bleach fluorescing species concentration. The bleach constant is a measure of the magnitude of the photobleaching that has been performed.

To model the reaction-diffusion of the species inside the bounded domain, we assume that a reaction of the form

$$F + B \leftrightarrow C \qquad (3)$$

describes the reaction of free diffusing molecules with binding sites where $F$ represents free species, $B$ represents vacant and immobile binding sites and $C$ represent bound complexes. Free molecules bind and unbind with rate constants $k_b$ and $k_u$, respectively. The average time for diffusion between binding events is $t_d=1/k_b$, while the average residency time of moving molecules in bound form is $t_r=1/k_u$. Unlike the $C$ species which remain fluorescent even when binding reaction occurs, $B$ are always non-fluorescent. Photobleaching is performed on a fluorescent population that has reached a uniform-steady state distribution. The bound population is considered to be immobile although this assumption may not be always correct for some complexes (Braga et al., 2007). By contrast, the unbound molecules are expected to diffuse freely. Furthermore, the distribution of the binding sites is considered to be homogeneous and remains constant during the fluorescence recovery due to their immobility.

Based on the above assumptions, Eq.3 is expressed mathematically as a system of reaction-diffusion equations (Sprague & McNally, 2005, Sprague et al., 2004, Tsibidis & Ripoll, 2008):

$$\frac{\partial f}{\partial t} = D_f \nabla_r^2 f - K_b f + k_u c$$
$$\frac{\partial c}{\partial t} = K_b f - k_u c$$
(4)

where $f$ and $c$ represent the concentration of fluorescent $F$ and $C$, respectively, $D_f$ is the diffusion coefficient of free population $F$ and $K_b \equiv k_b Bo$ (i.e. $Bo$ reflects the constant distribution of the binding sites). The concentration of $f$ and $c$ is measured in moles/$\mu m^2$. The subscript $r$ in the Laplacian operator indicates that all axial terms have been removed from the equation due to the two-dimensional character of the process. Given the Gaussian form of the bleaching profile, the initial conditions are

$$f(r,t=0) = \frac{1}{1+\gamma} C_i \exp(-K \exp(-2r^2/r_o^2))$$
$$c(r,t=0) = \frac{\gamma}{1+\gamma} C_i \exp(-K \exp(-2r^2/r_o^2))$$
(5)

where $\gamma = K_b/k_u$, and $F_o \equiv 1/(1+\gamma)$ and $C_o \equiv \gamma/(1+\gamma)$ represent the proportions of fluorescent free and bound populations, respectively. Homogenous Neumann boundary conditions are imposed at $r=b$ which imply there is no flux in and out of the boundary of the domain (Braga et al., 2007, Carrero et al., 2004a, Hinow et al., 2006, Sprague et al., 2006, Tsibidis & Ripoll, 2008):

$$\frac{\partial f(r=b, t \geq 0)}{\partial r} = \frac{\partial c(r=b, t \geq 0)}{\partial r} = 0$$
(6)

We now consider the behaviour of the system right after the bleaching: If diffusion of molecules is an instantaneous process, the concentration of free molecules (*f*) rapidly reaches an equilibrium state within a time interval which is short compared to the timescale of the FRAP experiment. Hence, only the concentration of fluorescent molecules that are bound during the bleaching phase (*c*) has a time-dependence on the time-scale of the FRAP experiment. Therefore, the total number of free fluorescent molecules inside the circle of radius *b* is always $\pi b^2 f$. Prior to bleaching, this simply corresponds to the total number of free molecules, whereas after bleaching, this product represents the number of free molecules that have not been bleached. Hence, one can write

$$\int_0^b dr\, 2\pi r F_o C_i \exp(-K \exp(-2r^2/r_o^2)) = \pi b^2 f$$
(7)

and thus, the post-bleach concentration of free species for *t>0* is

$$f = F_O C_i \left[ 1 + \sum_{n=1}^{\infty} \frac{(-K)^n}{2n!n} \left( \frac{r_o}{b} \right)^2 \left( 1 - \exp(-2nb^2/r_o^2) \right) \right] \quad (8)$$

To quantify FRAP, we consider a circular region of radius $w$ as a region of interest (Fig.1). The concentration of the fluorescent bound molecules inside the circular spot can be calculated from the second of Eqs.4 and Eq.8

$$c(t) = C_O C_i + \frac{C_O C_i}{2} \left( \frac{r_o}{b} \right)^2 \sum_{n=1}^{\infty} \frac{(-K)^n}{n!n} \left( 1 - \exp(-2n(b/r_o)^2) \right) +$$
$$C_O C_i \left[ \frac{1}{2} \left( \frac{r_o}{w} \right)^2 \sum_{n=1}^{\infty} \frac{(-K)^n}{n!n} \left( 1 - \exp(-2n(w/r_o)^2) \right) - \right.$$
$$\left. \frac{1}{2} \left( \frac{r_o}{b} \right)^2 \sum_{n=1}^{\infty} \frac{(-K)^n}{n!n} \left( 1 - \exp(-2n(b/r_o)^2) \right) \right] \exp(-k_u t) \quad (9)$$

As a result, the normalised (i.e. with respect to the pre-bleach value) value of total fluorescence is provided by the average of the sum of $f$ and $c(t)$

$$F(t) = 1 + \frac{1}{2} \left( \frac{r_o}{b} \right)^2 \sum_{n=1}^{\infty} \frac{(-K)^n}{n!n} \left( 1 - \exp(-2n(b/r_o)^2) \right) +$$
$$\frac{\gamma}{1+\gamma} \left[ \frac{1}{2} \left( \frac{r_o}{w} \right)^2 \sum_{n=1}^{\infty} \frac{(-K)^n}{n!n} \left( 1 - \exp(-2n(w/r_o)^2) \right) - \right.$$
$$\left. \frac{1}{2} \left( \frac{r_o}{b} \right)^2 \sum_{n=1}^{\infty} \frac{(-K)^n}{n!n} \left( 1 - \exp(-2n(b/r_o)^2) \right) \right] \exp(-k_u t) \quad (10)$$

A similar procedure can be followed (see Appendix) to derive the normalised fluorescence recovery curve for a circular disk bleaching profile (uniform bleaching) leading to the following formula

$$F(t) = \left[ 1 - \frac{\gamma}{1+\gamma} \exp(-k_u t) \right] \left( 1 - \left( \frac{w}{b} \right)^2 \right) \quad (11)$$

Eq.11 represents a modification to the FRAP expression derived in a previous work (Sprague et al., 2004), when molecular transport takes place in a bounded region.

By contrast, Eq.10 leads to the following expression in the infinite domain case ($b \to \infty$)

$$F(t) = 1 + \frac{\gamma}{1+\gamma}\left[\frac{1}{2}\left(\frac{r_o}{w}\right)^2 \sum_{n=1}^{\infty} \frac{(-K)^n}{n!n}\left(1 - \exp(-2n(w/r_o)^2)\right)\right]\exp(-k_u t) \quad (12)$$

Eqs.10-12 demonstrate the independence of the fluorescence recovery curves on the diffusion coefficient for fast moving species. This is a reasonable result because diffusion occurs so rapidly that it is not detected in the FRAP recovery. As a result, the diffusion coefficient which is a measure of the speed of molecular movement should not appear in a FRAP expression. By contrast, fluorescence recovery is solely attributed to binding kinetics.

## 3  Results and Discussion

### 3.1  Experimental validation

The approach described in the previous section can facilitate a quantitative interpretation of rapidly moving molecules that undergo binding-unbinding events. To test the significance of the proposed theoretical framework, the method was applied to investigate dynamics of nuclear proteins. Firstly, HeLa cells were transfected with GFP-actin and photobleaching was performed inside the cell nucleus. The cell nucleus is assumed to be a circular bounded region with a membrane at $b=6\mu m$. Nuclear actin is present in both globular and filamentous forms. Globular actin is free to diffuse whereas actin in filamentous form is assumed to be immobile and homogeneously distributed in the nucleus. Previous investigations showed that nuclear GFP-actin FRAP data exhibit a biphasic behaviour: a fast diffusion phase and a slow turnover phase (Carrero et al., 2004a, McDonald et al., 2006). A set of FRAP experiments produced average values for the fluorescence intensity (*green* line in Fig.2A) and fitting of the data was performed with Eq.11, a formula based on a previous work (Sprague et al., 2004) that contains a correction related to the presence of the boundary at $b$. The fitting is illustrated (*blue* solid line in Fig.2A) which yields $K_b=0.0034 sec^{-1}$ and $k_u=0.0183 sec^{-1}$.

We applied our model to obtain quantitative information about the slow phase that corresponds to binding kinetics. We obtained the best-fit value of the model parameters ($r_o$, $K_b$, $k_u$, $K$) by minimising the sum of the squared residuals $S$ between experimental data and the model. The parameters $r_o$, $K$ are the optical characteristics and they are related to the amount of bleaching applied to the specimen and the spread of the irradiation. By contrast, the other two parameters characterise the interaction of proteins with binding traps and they help to obtain the average binding time per site, $1/K_b$, and the average time for diffusion to the next site, $1/k_u$. Our proposed model allows the calculation of the binding-unbinding rate constants, the bleach constant and the radial resolution of the laser beam. The fitting of the experimental data with Eq.10 predicts $K_b=0.0057 sec^{-1}$ and $k_u=0.0225 sec^{-1}$, $r_o=1.269\mu m$, $K=2.9463$ (*magenda* line in Fig.2A). The amount of proteins free to diffuse (i.e. $K_b/(K_b+k_u)$) is 0.80% whereas only 20% is bound. Our model gives a slightly smaller $S$ ($S_{our\_model}=0.000403$ compared to $S_{sprague}=0.00045$), and one can expect the rate constants to be more precise because a more realistic bleaching profile is considered.

A second set of experimental data (*green* line in Fig.2B) that illustrate FRAP of GFP-β-actin was analysed (Fig.2B) based on results from (McDonald et al., 2006). Our model (*magenda* line in Fig.2B) yielded $K_b=0.0085 sec^{-1}$, $k_u=0.0336 sec^{-1}$, $r_o=1.6015\mu m$, $K=3.1808$, and $S_{our\_model}=0.0019$ while Eq.11 yielded a less accurate ($S_{sprague}=0.0036$) fitting and $K_b=0.0035 sec^{-1}$, $k_u=0.0159 sec^{-1}$.

## 3.2 Influence of laser beam parameters on the dynamic properties of the diffusing species

In the following, we will investigate the influence of the laser beam characteristics on the dynamic properties of the rapidly diffusing molecules. To simulate FRAP, we set the values for the radial resolution and the radius of the region of interest to $r_o=1.2\mu m$ and $w=1\mu m$, respectively. Movement of the molecules are assumed to be constrained inside a circular region of a radius equal to $b=6\mu m$. Figure 3A illustrates the initial fluorescence profile for four values of the bleaching constant (i.e. $K=1,2,4,7$) which are tested against the initial fluorescence resulted from the step function-like bleaching beam. Different bleaching strength and spread of the fluorescence depletion lead to different patterns of the bleached region (Fig.3A).

To demonstrate that binding reaction dominates the process, the results derived from the analytical solution (Eq.10) are compared to the results from a numerical solution of Eqs.4 (with the requirement that the initial (Eq.5) and boundary conditions (Eq.6) are satisfied). We have truncated the infinite series after $n_{max}=20$ terms (our calculations show the inclusion of additional terms does not contribute significantly to the value of fluorescence) and we set $K_b=10^{-4} sec^{-1}$ and $k_u=10^{-4} sec^{-1}$. The good agreement between the data from the two methods (i.e. sum of squared residuals $S<0.01$) demonstrates the adequacy of Eq.10 for describing processes characterised by instantaneous diffusion. In Figure 3B, the FRAP curves are plotted for four values of the bleaching constant while the *stars* correspond to data derived from a numerical solution of Eq.4. As expected, an increase of the bleaching constant leads to a greater bleaching effect. The comparison of the intensity curves for a Gaussian (Eq.10) and a uniform bleaching profile (Eq.11) suggests that recovery occurs at a lower speed in the latter case. Results from a more rigorous and realistic model (Eq.10) are different from the approximating solution expressed by Eq.11. As shown in the previous section, the absence of the bleach constant and the laser beam resolution from the analytical expressions of simpler models (Sprague et al., 2006, Sprague et al., 2004) leads to less accurate values for the rate constants.

As mentioned above, while the value of the bleaching constant determines the degree of bleaching performed on a particular region of the sample, the laser beam resolution $r_o$ is related to the spread of bleaching. Larger values of $r_o$ suggest a larger photobleached area (Figure 4A). Thus, the total pre-bleach amount (or equally the concentration at large time values) of fluorescent material inside the bounded region reduces to smaller values as $r_o$ increases (Figure 4B).

The role of the size of the bounded domain to the speed of the recovery is emphasised by the comparison of the FRAP curves for three barrier sizes (i.e. $b=4, 10$ and $20\mu m$). In Figure 5A, all of the curves correspond to the same pair of reaction rates ($K_b=10^{-4} sec^{-1}$ and $k_u=10^{-4} sec^{-1}$) and the bigger the barrier size, the better fitting with the infinite domain curve expressed by Eq.11. Previous studies have shown (Carrero et al., 2004a, Tsibidis & Ripoll, 2008) that unless the molecules diffuse rapidly (i.e. the rate constant combination lies in the reaction dominant regime), the initial phase of the fluorescence recovery is independent of the domain boundary. As a result, at early timepoints, expressions which are valid for the infinite domain can be perfectly used to fit experimental data that describe fluorescence recovery for bounded domains producing the same quantitative information about the interaction with immobile structures. By contrast, for fast diffusing molecules, the presence of a boundary appears to influence the form of the recovery curve in the early stages (Fig.5A).

Previous studies have also shown that for a uniform bleaching profile (Sprague et al., 2004, Sprague et al., 2006), FRAP curves are independent of the size of the bleached region for rapidly moving species. This property is used to determine whether fluorescence recovery was attributed entirely to reaction. By contrast, our approach suggests there is a dependence of the fluorescence recovery expression on the size of the region of interest $w$ (Eq.10 and Eq.11) and therefore this criterion is no longer valid. In Fig.5B we varied both $w$ and the laser beam radial resolution $r_o$ while keeping always their ratio constant and we observed that as we increased the two variables, fluorescence was reaching the maximum value faster.

# 4 Conclusions

A technique has been presented to assess molecular transport and interactions with other components inside domains when diffusion occurs much faster than binding and the timescale of the FRAP experiment. Our model attempted to expand and complement previous investigations by incorporating additional parameters related to the laser beam characteristics of the CSLM. Analytical expressions were derived that were applied to investigate protein kinetics inside the cell nucleus. The model aims not only to facilitate our understanding of molecular interaction by yielding the association and dissociation rates of the interaction but also to provide useful insights into the general principles of species kinetics.

**Acknowledgements**

The author would like to acknowledge financial support from the EU Coordination Action Project ENOC 022496.

**Appendix**

**Derivation of fluorescence recovery curve for a circular disk bleaching profile (uniform bleaching)**

If diffusion of molecules is a very fast process, the concentration of free molecules (*f*) rapidly reaches an equilibrium state within a time interval which is short compared to the timescale of the FRAP experiment. Considering a uniform distribution of fluorescence material inside the domain of radius $b$ (Fig.1), the total number of free fluorescent molecules at the equilibrium state should be given by

$$N1 \equiv \int_0^b dr\, 2\pi r F_o C_i \quad \text{(A.1)}$$

On the other hand, for a uniform circular bleaching profile, fluorescence inside a bleached spot of radius $w$ vanishes while it remains constant outside the spot. Hence, the total number of free fluorescent molecules is given by

$$N2 \equiv \int_w^b dr\, 2\pi r f \qquad (A.2)$$

Conservation of the number of fluorescent molecules implies that *N1=N2* which yields the following post-bleach concentration of free fluorophores for *t>0*

$$f = C_i F_O \left(1 - \left(\frac{w}{b}\right)^2\right) = C_i \frac{1}{1+\gamma}\left(1-\left(\frac{w}{b}\right)^2\right) \qquad (A.3)$$

If we substitute Eq.A.3 into the second equation of Eq.4 and we solve the differential equation, we obtain the concentration of the fluorescent bound molecules *c(t)* (given that *c(t=0)=0*)

$$c(t) = C_i C_O [1-\exp(-k_u t)]\left(1-\left(\frac{w}{b}\right)^2\right) = C_i \frac{\gamma}{1+\gamma}[1-\exp(-k_u t)]\left(1-\left(\frac{w}{b}\right)^2\right) \qquad (A.4)$$

By adding Eq.A.3 and Eq.A.4, we obtain the normalised (i.e. with respect to the pre-bleach value) fluorescence recovery expression

$$F(t) = \left[1 - \frac{\gamma}{1+\gamma}\exp(-k_u t)\right]\left(1-\left(\frac{w}{b}\right)^2\right) \qquad (A.5)$$

which represents the modified version of a formula derived previously (Sprague et al., 2004) when molecule transport takes place in a bounded region.

**List of Figures**

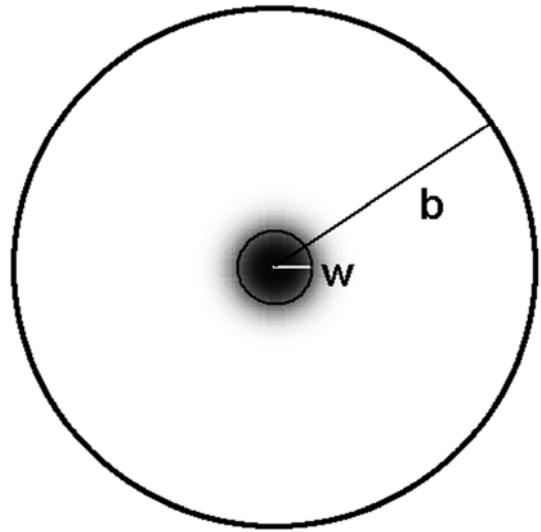

Figure 1: Bleaching profile and initial intensity. A circular region of radius *w* is used as a region of interest to quantify the fluorescence recovery after bleaching in a bounded domain of diameter *2b* after a laser beam with a Gaussian bleaching profile was used.

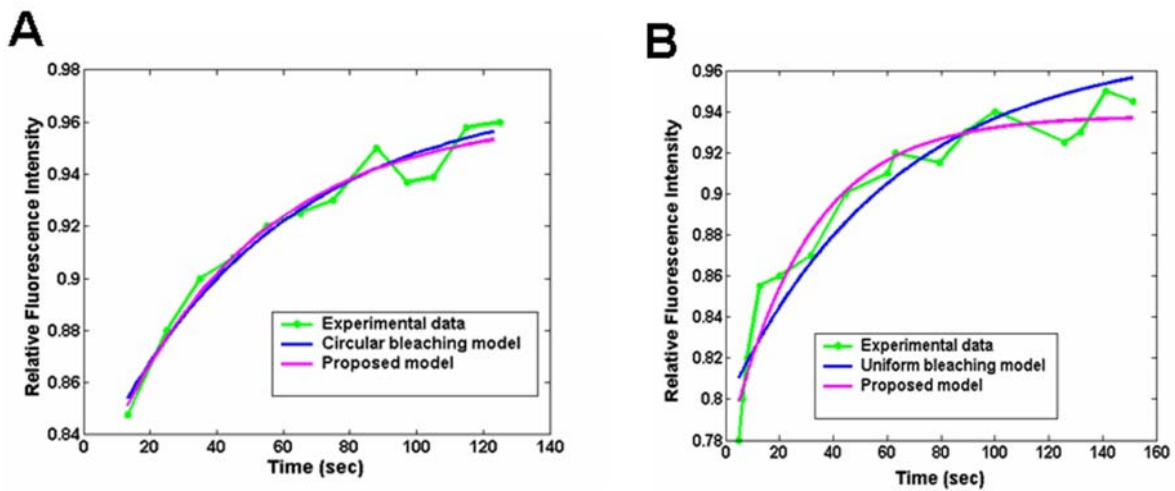

Figure 2: Fitting of experimental data that describe FRAP of GFP-β-actin in HeLa cells (*green* line) with the proposed (*magenda* line) and a model based on a circular bleaching profile (*blue* line). Experimental data have been obtained from (A) (Carrero et al., 2004a), and (B) (McDonald et al., 2006).

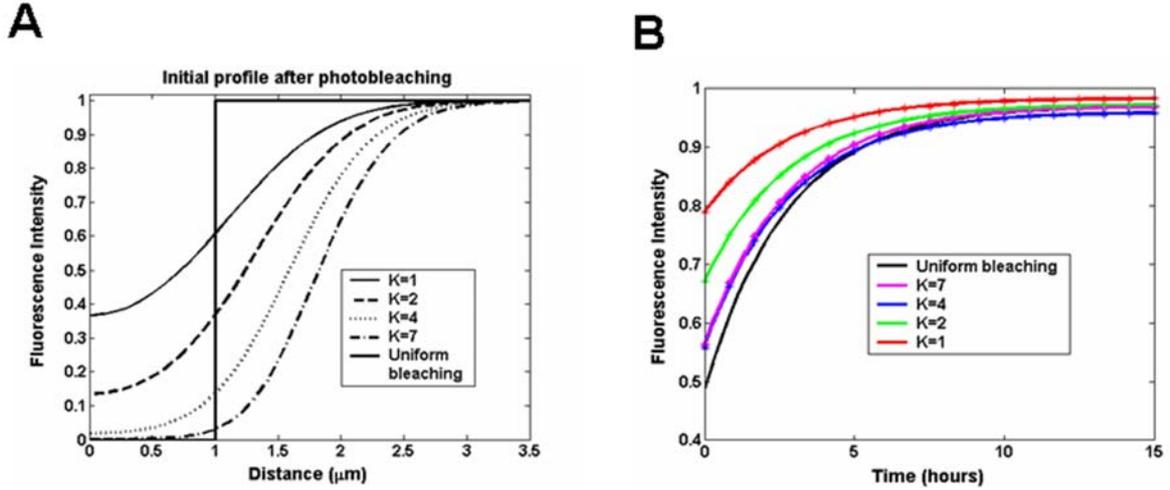

Figure 3: (A) Initial fluorescence profile after photobleaching was performed on a cell with a Gaussian laser bleaching beam of radial resolution $r_o$=1.2μm. The dependence of the curve on the bleaching constant $K$ is demonstrated. The *thick solid* line represents the initial profile for a uniform bleaching. (B) FRAP curves are plotted for four values of $K$ (for $w$=1μm, $K_b$=10$^{-4}$ sec$^{-1}$, $k_u$=10$^{-4}$sec$^{-1}$, $r_o$=1.2μm). The *dots* represent the data derived from a numerical solution of Eq.4, the *black* line corresponds to the FRAP curve derived from the uniform bleaching model (Eq.11).

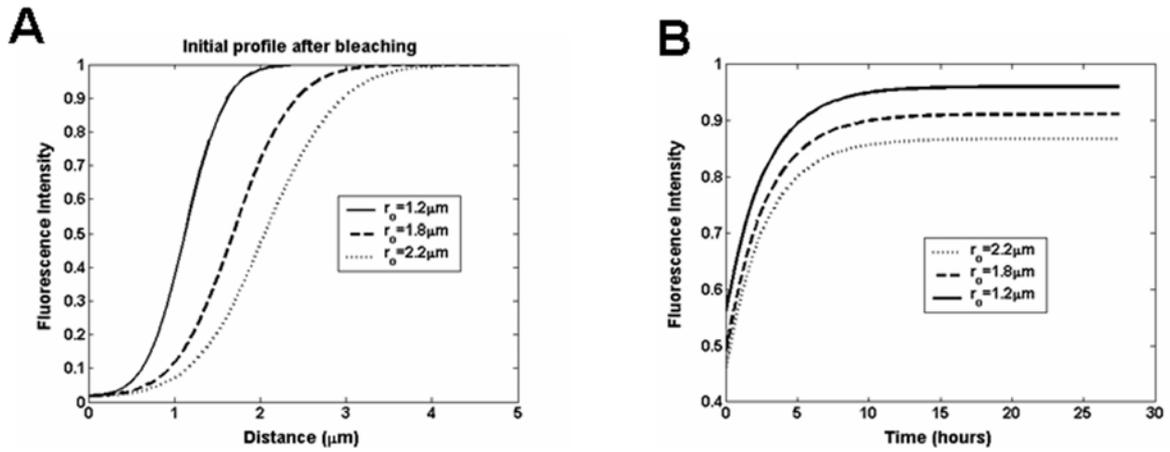

Figure 4: (A) The dependence of the initial intensity on $r_o$ is demonstrated. (B) FRAP curves are plotted for various values of $r_o$. The parameter values used in the simulation are: $K$=4, $w$=1μm, $K_b$=10$^{-4}$sec$^{-1}$, $k_u$=10$^{-4}$ sec$^{-1}$.

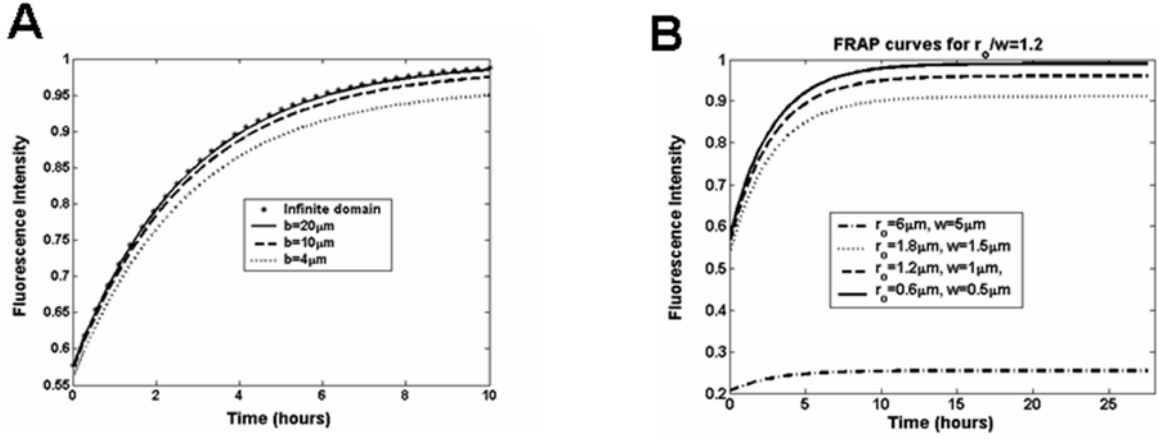

Figure 5: FRAP curves that indicate the dependence of the recovery process on (A) the barrier size $b$ (for $K=4$, $w=1\mu m$, $K_b=10^{-4}sec^{-1}$, $k_u=10^{-4}sec^{-1}$, $r_o=1.2\mu m$), and (B) the Gaussian radial resolution of the laser beam $r_o$ and size of the region of interest $w$ (for $K=4$, $w=1\mu m$, $K_b=10^{-4}sec^{-1}$, $k_u=10^{-4}sec^{-1}$, $b=6\mu m$).